\documentclass[10pt,letterpaper]{article}

\usepackage{graphicx}
\usepackage{caption}
\usepackage{subcaption}
\usepackage{cogsci}
\usepackage{pslatex}
\usepackage{apacite}
\usepackage{amsmath}
\usepackage{amssymb}
\usepackage{wrapfig}
\usepackage[export]{adjustbox}

\title{An Experimental Protocol to Derive and Validate a Quantum Model of Decision-Making}
 
\author{
        {\large \bf Lauren Fell (l3.fell@qut.edu.au)} \AND 
        {\large \bf Shahram Dehdashti (shahram.dehdashti@qut.edu.au)} \AND
        {\large \bf Peter Bruza (p.bruza@qut.edu.au)} \AND
        {\large \bf Catarina Moreira (catarina.pintomoreira@qut.edu.au)} \AND
        School of Information Systems, Queensland University of Technology\\
        Brisbane, Australia.
}

\begin{document}

\maketitle

\begin{abstract}
This study utilises an experiment famous in quantum physics, the Stern-Gerlach experiment, to inform the structure of an experimental protocol from which a quantum cognitive decision model can be developed. The 'quantumness' of this model is tested by computing a discrete quasi-probabilistic Wigner function.
Based on theory from quantum physics, our hypothesis is that the Stern-Gerlach protocol will admit negative values in the Wigner function, thus signalling that the cognitive decision model is quantum. 
A crowdsourced experiment of two images was used to collect decisions around three questions related to image trustworthiness. The resultant data was used to instantiate the quantum model and compute the Wigner function.
Negative values in the Wigner functions of both images were encountered, thus substantiating our hypothesis. Findings also revealed that the quantum cognitive model was a more accurate predictor of decisions when compared to predictions computed using Bayes' rule. 

\end{abstract}
\section{Introduction}
\indent 
A generally accepted notion is that we can approximately access cognitive states through questioning and observation, and whilst this measurement is not deemed to be perfect, it is a standard means of experimental practice in the psychological discipline. This notion relies on the fact that these internal states hold distinct values and by measuring them, we are merely attempting to record what is already there. Often, probabilistic outcomes of these measures appear to be illogical and do not follow the laws of classical probability, for example, cognitive biases identified in decision-making \cite{tversky1974judgment}. Quantum cognition has emerged as an alternative means of analysing probabilistic outcomes that do not follow these classical laws. Its potential derives from an alternative probability which has successfully been used to address human decision making considered paradoxical, generate non-reductive understandings of human conceptual processing, and provide new understandings of perception and human memory \cite{bruza:wang:busemeyer:2015,busemeyer:bruza:2012}. 
The present paper extends current approaches to quantum modelling by means of two new aspects: 
1) the Stern-Gerlach experiment, to inform an experimental protocol from which a complex Hilbert space model can be constructed and 2) the discrete Wigner function to perform a check on 'quantumness' of cognitive systems being modelled. 
\begin{figure}
\centering
  \includegraphics[width=8cm]{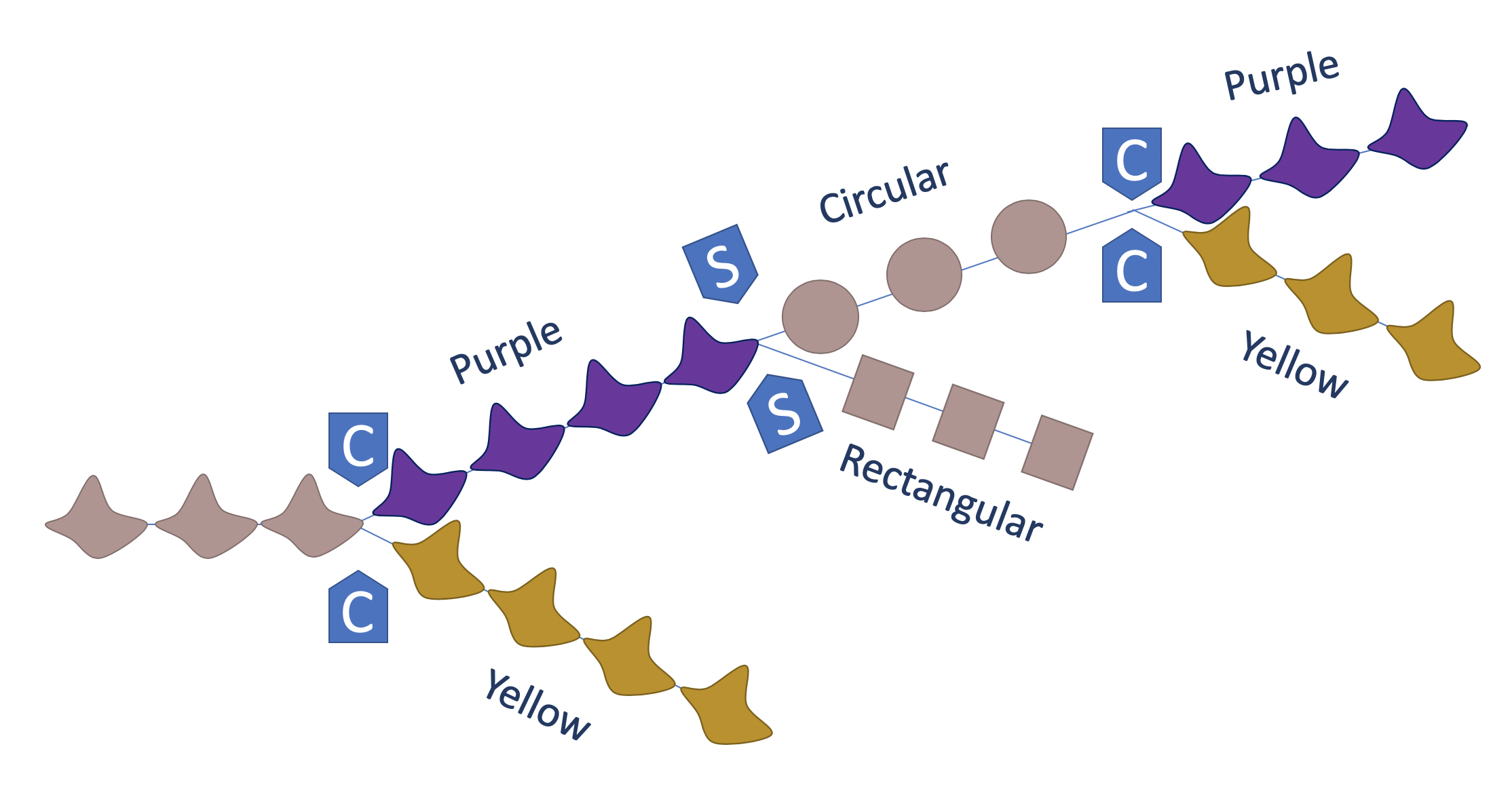}
  \caption{S.G. Setup using colour-type measurements (C) and shape-type measurements (S) in place of spin measurements at different orientations.}
\label{SGSetup}
\end{figure}
\subsection{The Stern-Gerlach Experiment}
 The  Stern-Gerlach (S.G.) experiment \cite{sakurai1995modern}  takes a beam of particles (for example, silver atoms) and observes their spin using a device that creates an electromagnetic field (S.G. device). This device can be placed at different orientations to observe spins at associated orientations. Due to the fact that a particle's spin is a complex concept to describe, we will substitute this property with colour and shape in our description of the S.G. experiment in the interests of clarity.  For the following, we will describe the experiment as having two orientations of S.G. devices: one oriented one way to measure one type of spin (we will call this a 'colour-type' measure differentiating purple from yellow), and another oriented orthogonal to the first to measure a second type of spin (we will call this a 'shape-type' measure differentiating circular from rectangular). Each enable measurement of separate aspects of the same object (two orientations of spin, or colour and shape in our analogy).\\
\indent The experiment involves a beam of atoms hitting an S.G. device which splits it into two separate beams: one purple beam and another yellow beam. A second S.G. device of the same orientation as the first (colour-type) is placed in the path of the purple beam. This time, the beam does not split, and only one beam of purple atoms come out, as one would expect (i.e. we assume no yellow atoms to have entered it due to the separation performed by the first S.G. device). We then place a shape-type S.G. device in-between the two colour-type S.G. devices, which splits the first purple beam into circular and rectangular atoms before hitting the second colour-type device (see figure \ref{SGSetup}). 
\begin{figure*}
\vspace{-1.5 cm}
\begin{center}
  \includegraphics[width=16cm]{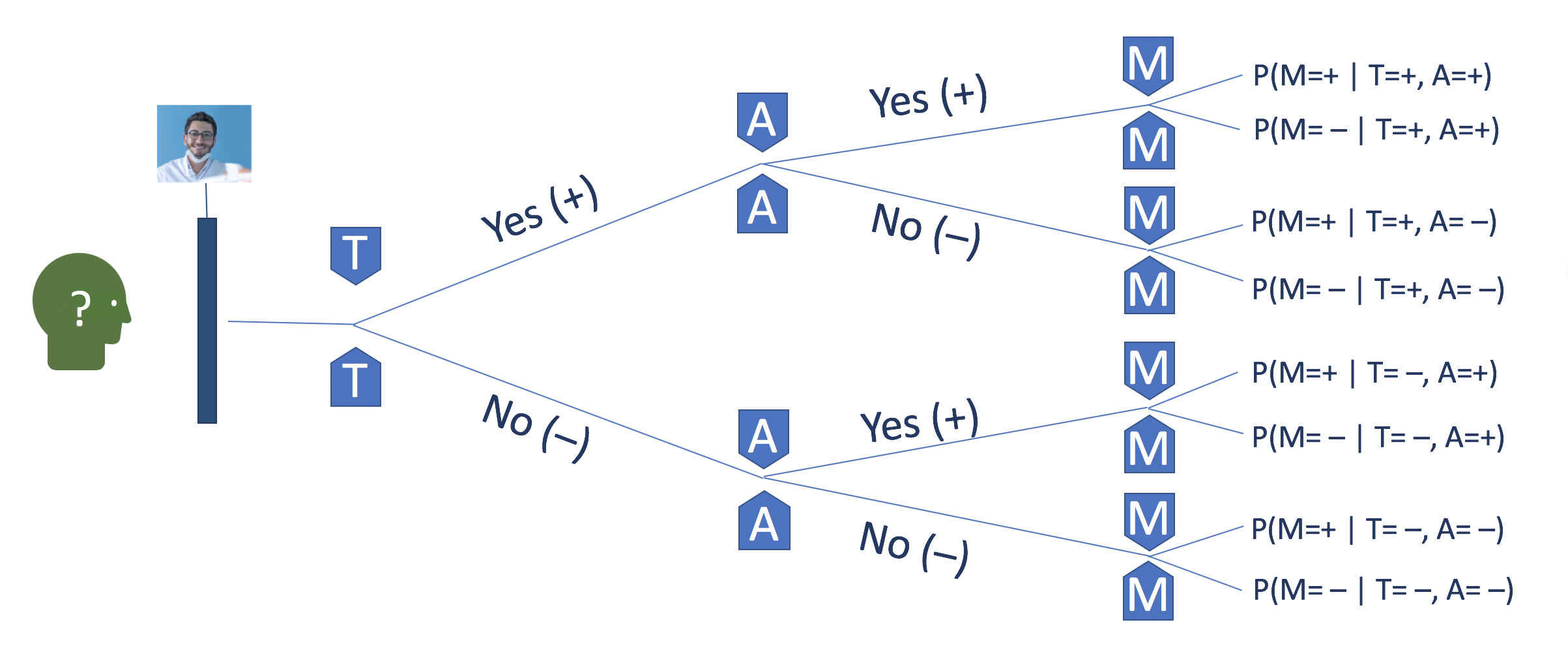}
  \caption{ A cognitive analogue to a S.G. experiment where Trustworthiness (T) is asked first, followed by Attractiveness (A), then Manipulated (M).}
  \label{fig1}
\end{center}
\end{figure*}

\indent In a classical system, we would expect only one beam to emit from the second colour-type device, as, again, we would assume that only purple atoms were sent that way from the first device. A quantum system, however, does not work in this way. In a quantum system, the second colour-type device will emit two beams after the original beam of purple atoms has passed through the shape-type device, as illustrated in figure \ref{SGSetup}. This is because the shape-type measurement has destroyed the first measurement of colour, essentially resetting it. This can only happen if a particle's properties are not simply observed in a predefined definite state, but are determined at the point of measurement.
\subsection{A Cognitive Analogue}
This concept can be applied to cognitive measures, however, in a slightly more complex way. Due to the potential of memory effects inherent in repeating a question in a string of only three total questions, we utilise a more complex version of the S.G. experiment, where a third question is introduced. The general concept, however, remains the same. When one considers a question or makes a decision, they may simply be accessing an internal state predetermined by a range of variables such as past experience, knowledge, predisposition, values, what they had for lunch that day, etc. On the other hand, they may be creating the state only at the point of measurement (i.e. considering a question or making a decision). To place this in the context of the S.G. experiment, consider three questions asked after presentation of an image: Do you feel a sense of trust when viewing this (T), do you feel that the person in this image is attractive (A), and do you feel that the image may have been manipulated (M). Taking a classical position, one could describe this system in the following way: A person views an image and this event interacts with internal variables to create a variety of probable judgments of this image, including judgments of trust, attractiveness and manipulation. A person then considers the sequential questions of trust, attractiveness and manipulation, each time taking an internal measurement of the predefined values that each of these hold in the person's internal state. On the contrary, taking a quantum position would instead describe no definite judgments to be formed at the point of viewing the image, but only at the point of considering each question. This view would also posit that each question would destroy the measurement of the prior question, in the same way the shape-type S.G. device did in our above example. \\
\indent 
This article presents an experimental protocol that is analogous to the S.G. experiment in order to derive a quantum model of decision making. For this purpose a complex Hilbert space is used.

\subsection{Derivation of a Quantum Model of Decision Making from the S.G. device}

As described above, the basic idea behind the model is to translate the S.G. device into cognitive science by way of analogy; human subjects correspond to silver atoms and questions correspond to S.G. devices. 
As a running example we will use an image trustworthiness task whereby subjects are asked whether they trust (T) an image, whether they find the subject of the image attractive (A) and whether they deem the image to be manipulated (M) e.g., photoshopped.
The particular order of questions is determined by the order of the devices in the  S.G. device as depicted in figure~\ref{fig1}.
 
The derivation of the quantum model corresponding to the S.G. device comprises two steps:
In the first step, a complex Hilbert space model  with states and operators is constructed.  In the second step, a criterion for checking the `quantumness' of the model is applied by using a discrete Wigner function. 

The cognitive decision space is modelled by means of a complex Hilbert space model (HSM), i.e., a complex vector space, equipped with an inner product with a positive definite metric \cite{sakurai1995modern}. Any yes/no outcome of a specific question $X$, is denoted by a ket $|X,\pm\rangle$ using the Dirac notation, where +/- respectively denotes a yes/no outcome :
\begin{eqnarray}\nonumber
|X\rangle=\alpha |X,+\rangle+\beta |X,-\rangle, \hspace{.5cm} \alpha, \beta \in \mathbb{C}
\end{eqnarray}
in which $|\alpha|^{2}$ and $|\beta|^{2}$, based on the Born rule, give the probability of observing the positive and negative answers. 
In addition, outcomes are orthogonal, $\langle X,\pm | X, \mp\rangle = 0$ and probabilities are normalized, $|\alpha|^{2}+|\beta|^{2}=1$.
Also, an observable is defined as a Hermitian operator. Without going into the technical details, a Hermitian operator $\hat{A}$ is a special type of matrix where the eigenstates correspond to outcomes that are observed, and the corresponding eigenvalue relates to the probability of observing that outcome.
\indent 
The Pauli matrices $\sigma_{i}$, $i=1,2,3$,
\begin{eqnarray}\nonumber
\sigma_{x}=\left(
\begin{array}{cc}
    0 & 1 \\
    1 & 0
\end{array}
\right),\hspace{.3cm}
\sigma_{y}=\left(
\begin{array}{cc}
    0 & -i \\
    i & 0
\end{array}
\right), \hspace{.3cm}
\sigma_{z}=\left(
\begin{array}{cc}
    1 & 0 \\
    0 & -1
\end{array}
\right),
\end{eqnarray}
along with the identity matrix $I_{2\times 2}$, form an orthogonal basis for the complex Hilbert space of all $2 \times 2$ matrices.
As a consequence, any operator such as $\hat{A}$ can be expressed by $\hat{A}=a_{0}I+\sum_{i=1}^{3} a_{i}\sigma_{i}$,
 in which $a_{i}\in \mathbb{R}$ with $i=0,1,2,3$.  
\subsubsection*{Steps to construct a complex HSM}
\indent Based on the preceding formalism, the following steps are used to derive the quantum model from the S.G. depicted in figure~\ref{fig1}:\\ 
\indent  1) A quantum state is defined by the first question $T$ based on relative frequencies of yes/no outcomes sampled from the experimental data,
$|T \rangle=\sqrt{P_{t}(+)}|T,+\rangle+
\sqrt[]{P_{t}(-)}|T,-\rangle$,
 in which $ P_{t}(+)$ and $ P_{t}(-)$ are respectively probability of finding positive and negative responses to the question $T$. Also, we can consistently define the projection or filtering-type quantum cognitive  operator $\hat{\pi}_{t}(\pm)=|T,\pm\rangle\langle T,\pm|$, so that 
 $ P_{t}(\pm)=\langle 
 T|\hat{\pi}_{t}(\pm)|T\rangle $. The filtering-type operators 
 $\hat{\pi}_{t}(\pm)$ satisfy the 
 completeness relation, 
 $\hat{\pi}^{\dagger}_{t}(+) 
 \hat{\pi}_{t}(+)+\hat{\pi}^{\dagger}_{t}(-) \hat{\pi}_{t}(-)=I_{2\times 2}$,
 and the operator $\hat{T}$ is defined as : $\hat{T}= \hat{\pi}_{t}(+)-\hat{\pi}_{t}(-)
 = \sigma_{z}$, where $\sigma_{z}$ is the Pauli matrix in direction $z$.\\
\indent 2) In the second step, we   obtain the probability of finding positive and negative responses to the second question $A$ through the first question $T$.  
Hence, we can define the cognitive  state regarding a decision of  attractiveness $A$ in the basis of the state of trustfulness $T$, i.e.,
$|A, + \rangle =\cos \frac{\theta_{a}}{2}|T,+\rangle + \sin \frac{\theta_{a}}{2}|T,-\rangle$ and
$|A, - \rangle =\sin \frac{\theta_{a}}{2}|T,+\rangle - \cos \frac{\theta_{a}}{2}|T,-\rangle.  
$
 The filtering-type measurement operators 
 $\pi_{a}(\pm)$  can be written  as follows:
\begin{eqnarray*}
\hat{\pi}_{a}(\pm)= 
 \frac{1}{2}\Big[I_{2\times 2} 
\pm \sin \theta_{a}  \sigma_{x}
\pm \cos\theta_{a} \sigma_{z}\Big].
\end{eqnarray*}
Hence, the operator $\hat{A}$ is given by
\begin{eqnarray}\label{ope}
\hat{A}=\left[
\begin{array}{cc}
\cos \theta_{a}& \sin \theta_{a}\\
\sin \theta_{a} & -\cos \theta_{a}
\end{array}
\right]
\end{eqnarray}
in which $\theta_{a} $  characterizes a specific direction, which can be computed from the experimental data.  
By applying the Born rule, the conditional probabilities can be computed.
 For example, 
\begin{eqnarray}\label{eq3}
P(A=+|T=+)=|\pi_{a}(+)\pi_{t}(+)|T\rangle|^{2}
=P_{t}(
+)\cos^{2} \frac{\theta_{a}}{2},\label{eq8}
\end{eqnarray}
 \indent 3) A similar method to step 2) derives another filtering-type operator corresponding to the third question $M$, 
\begin{eqnarray}
\hat{\pi}_{m}(\pm)
= \frac{1}{2}\Big[I_{2\times 2}
\pm  \sin \theta_{m} \cos \phi_{m} \sigma_{x}
\pm\sin \theta_{m} \sin \phi_{m} \sigma_{y}\pm \cos\theta_{m} \sigma_{z}\Big]\nonumber
\end{eqnarray}
in which $\theta_{m} $ can be obtained, despite of the fact that we must have extra information for acquiring  $\phi_{m}$.
Note that the states of third question $M$ are defined as follows:
\begin{eqnarray}
    |M,+\rangle&=&\cos \frac{\theta_{m}}{2}|T,+\rangle+e^{i\phi_{m}}\sin \frac{\theta_{m}}{2}|T,-\rangle,\nonumber\\
    |M,-\rangle&=&e^{-i\phi_{m}}\sin \frac{\theta_{m}}{2}|T,+\rangle+\cos \frac{\theta_{m}}{2}|T,-\rangle.\nonumber
\end{eqnarray}
\indent 4) In the last step, probabilities of the third question $M$ are computed based in light of the outcomes from the second question $A$: 
\begin{eqnarray}\label{eqf3}
&P(M=+|A=+,T=+) 
=|\pi_{m}(+)\pi_{a}(+)\pi_{t}(+)|T\rangle|^{2}\nonumber\\
&=P_{t}(+)\cos^{2} \frac{\theta_{a}}{2}(
+)\Big(\cos^{2} \frac{\theta_{a}}{2}\cos^{2} \frac{\theta_{m}}{2}
+\sin^{2} \frac{\theta_{a}}{2} \sin^{2} \frac{\theta_{m}}{2}\nonumber\\
&\hspace{-4cm}+\frac{1}{2}\sin \theta_{a} \sin \theta_{m} \cos \phi_{m}\Big).
\end{eqnarray}
By using the previous equations, values of $\phi_{m}$ can be computed.
%
%
%

\subsubsection{Determining quantumness using the discrete Wigner distribution}

\begin{figure}[t]
\centering
  \includegraphics[width=8cm]{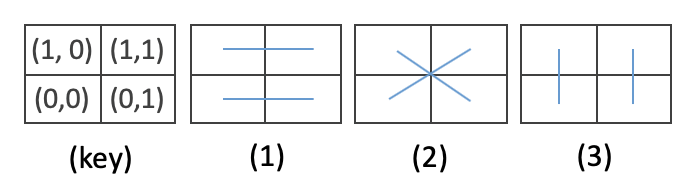}\\
  \caption{The striations of the $2\times 2$ phase space. Each point occupies a quadrant.
}\label{fig11}
\end{figure}

When we construct the Hilbert space structure of the cognitive  state and associated operators, we can examine the quantumness of the cognitive state. 
Quantum physics has a range of criteria for this. In this article we will employ one such criterium, namely the negative discrete Wigner function, where the negativity of the function can be interpreted as a signature of quantum interference.
In order to explain the discrete Wigner function, the continuous Wigner function is first introduced.
For a continuous phase space $(q,p)$, the continuous Wigner distribution is defined by
\begin{eqnarray}
W_{\Psi}(q,p)=\frac{1}{2\pi}\int_{-\infty}^{\infty}dx \langle q-\frac{x}{2}|x\rangle\langle x|q+\frac{x}{2} \rangle e^{ipx}.
\end{eqnarray}
Therefore, the expectation value of an arbitrary operator $\hat{X}$, by using the Wigner distribution, is given by
\begin{eqnarray}
\langle \hat{X}\rangle=Tr [\hat{\rho}\hat{X}]= \int \int dx\ dp W(x,p) \tilde{X} (x,p),
\end{eqnarray}
in which $\tilde{X}(x,p)$ is the average of a physical quantity over the phase space.
Assuming two arbitrary states $|\psi_{a}\rangle$ and $|\psi_{b}\rangle$, it can be verified that: 
\begin{eqnarray}
|\langle \psi_{a}|\psi_{b}\rangle|^{2}=Tr[\hat{\rho}_{a}\hat{\rho}_{b}]=\int dx\ dp W_{\psi_{a}}(x,p) W_{\psi_{b}}(x,p).
\end{eqnarray}
If we consider a situation in which two states are orthogonal, i.e.,
\begin{eqnarray}
\int dx\ dp W_{\psi_{a}}(x,p) W_{\psi_{b}}(x,p)=0,
\end{eqnarray}
at least in part of the region in phase space, one of  the above mentioned Wigner distributions has to be negative. 
The values in Wigner distributions still sum to 1 even when values happen to be negative, which is why Wigner distributions are termed ``quasi-probability" distributions. The negativeness of the Wigner distribution can be the result of the following two facts: Firstly, the accessibility of information for a system described by the quantum formalism is when the system is described by classical probability  \cite{goh2018geometry, vourdas2019probabilistic}. Secondly, the negativeness can be interpreted as quantum contextuality   \cite{huangnegativity,raussendorf2017contextuality,kocia2017discrete}.    \\
\indent The binary nature of the responses implies a discrete, rather than continuous phase space.
We apply a generalized version of the continuous Wigner function \cite{wootters1987wigner}. 
In fact, by defining a geometrical structure on the discrete phase space, such as parallel line, {\it etc.}, and using a Finite Field $\mathcal{F} _{n}$, a discrete Wigner distribution can be defined \cite{gibbons2004discrete, galvao2005discrete, PhysRevA.95.022340}. \\
\indent
Due to the fact that we have binary responses, the discrete phase space occupies a  $2 \times 2$ array of points where   $q$
runs along the horizontal axis and $p$ runs along the vertical
axis, as shown in figure \ref{fig11}.
We place the origin, $(q, p)=(0,0)$, at
the lower left-hand corner. We define a line $\lambda$ in the $2\times 2$ phase space as the set
of two points satisfying an equation of the form $aq+bp=c$, where
$a$, $b$, and $c$ are elements of $\mathbb{Z}_{2}$ ($\mathbb{Z}_{2}$ constraints numbers to binary 0s and 1s) where $a$ and $b$ cannot both equal zero. It has the following conditions: (i) given
any two distinct points, exactly one line contains both points;
(ii) given a point $\alpha$, if a line $\lambda$ does not contain $\alpha$, there is
exactly one line parallel to $\lambda$ that does contain it (iii) two lines
that are not parallel intersect in exactly one point. 
In the preceding conditions, two lines can be considered parallel if they can be represented by equations
having the same values for $a$ and $b$ but different values for $c$. In the case of a binary response, therefore, a line connecting $(0,0)$ and $(0,1)$ is parallel with the line that connects $(1,0)$ and $(1,1)$. Moreover, two equations $p+q=0$ and $p+q=1$, with $p, q \in \mathbb{Z}_{2}$, give the lines connecting points $(1,0)$ and $(0,1)$ and the parallel line connecting $(0,0)$ and $(1,1)$. Finally, the line  $(0,0)$ and $(1,0)$ is parallel with the line $(0,1)$ and $(1,1)$. Figure \ref{fig11} demonstrates
these striations in (1), (2), and (3) respectively. Note that the lines drawn in (2) are technically parallel based on the equation described above.
As is the case in the continuous phase space,
the integral of the Wigner function over the strip of phase space bounded by the lines
$aq+bp=c_{1}$ and $aq+bp=c_{2}$ is the probability that the operator
$a\hat{q} +b\hat{p}$ will take a value between $c_{1}$ and $c_{2}$ \cite{wootters1987wigner}, the discrete Wigner function has to satisfy the following equation:
\begin{eqnarray}
Tr\left(|\alpha_{i,j} \rangle \langle \alpha_{i,j}| \rho\right)=\sum_{\alpha \in \lambda_{i,j}} W_{\alpha},
\end{eqnarray}
in which $\rho$ is density matrix,
\begin{eqnarray}\label{eq7}
\hat{\rho}= \frac{1}{2}\left(I+\vec{r}\cdot \sigma
\right)=\frac{1}{2}\left(
\begin{array}{cc}
1+r_{z} & r_{x}-ir_{y}\\
 r_{x}+ir_{y} & 1-r_{z}
\end{array}
\right).
\end{eqnarray}
and
$|\alpha_{i,j}\rangle$ are three  mutually unbiased bases  for a two-dimensional Hilbert
space, with the following property: 
\begin{eqnarray}
|\langle \alpha_{i,j}|\alpha_{k,l}\rangle|^{2}=\frac{1}{2} \ \text{if } \ i \neq k,
\end{eqnarray}
where $i=1,2,3$     indexes the
mutually unbiased bases
and $j=1,2$  indexes the basis vector in each mutually unbiased bases, with the following 
condition
 $\langle \alpha_{i,j}|\alpha_{i,k}\rangle|^{2}=\delta_{j,k}$. Naturally, we can consider a one-to-one map between Pauli matrices $\sigma_{i}$ and striations $S_{i}$.
\section{Experiment}

\subsection{Participants}
Participants consisted of 300 members of the crowdsourcing platform Prolific, 187 of which were male, 110 female, and 3 who preferred not to disclose their gender. Participants were over 18 years and from a variety of countries across North America (39.7\%), Europe (32.3\%), UK (22.9\%), Australiasia (4.0\%) Middle East (0.7\%) and Asia (0.3\%).
Participants were randomly assigned to one of 4 conditions, each with 75 participants. All participants had been verified as proficient in English by Prolific. Remuneration was in the form of a small payment (\pounds.23), as per Prolific convention, and an informed consent page was presented to participants prior to commencement.
\subsection{Materials}
\begin{figure}
\centering
\begin{subfigure}{.5\columnwidth}
  \centering
  \includegraphics[width=4cm]{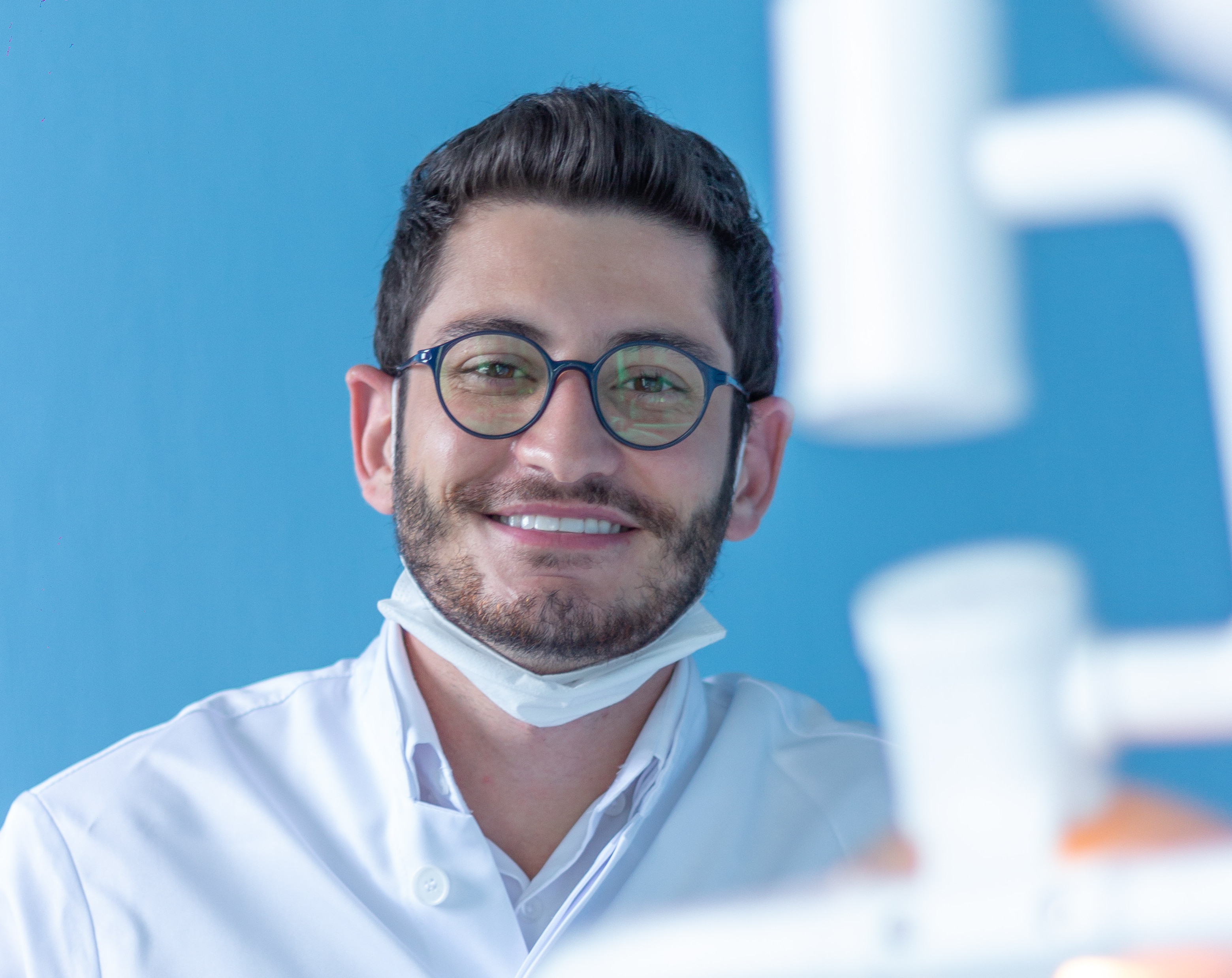}
  \caption{Image 1: Unedited.}
  \label{doctorundeditimage}
\end{subfigure}%
\begin{subfigure}{.5\columnwidth}
  \centering
  \includegraphics[width=4cm]{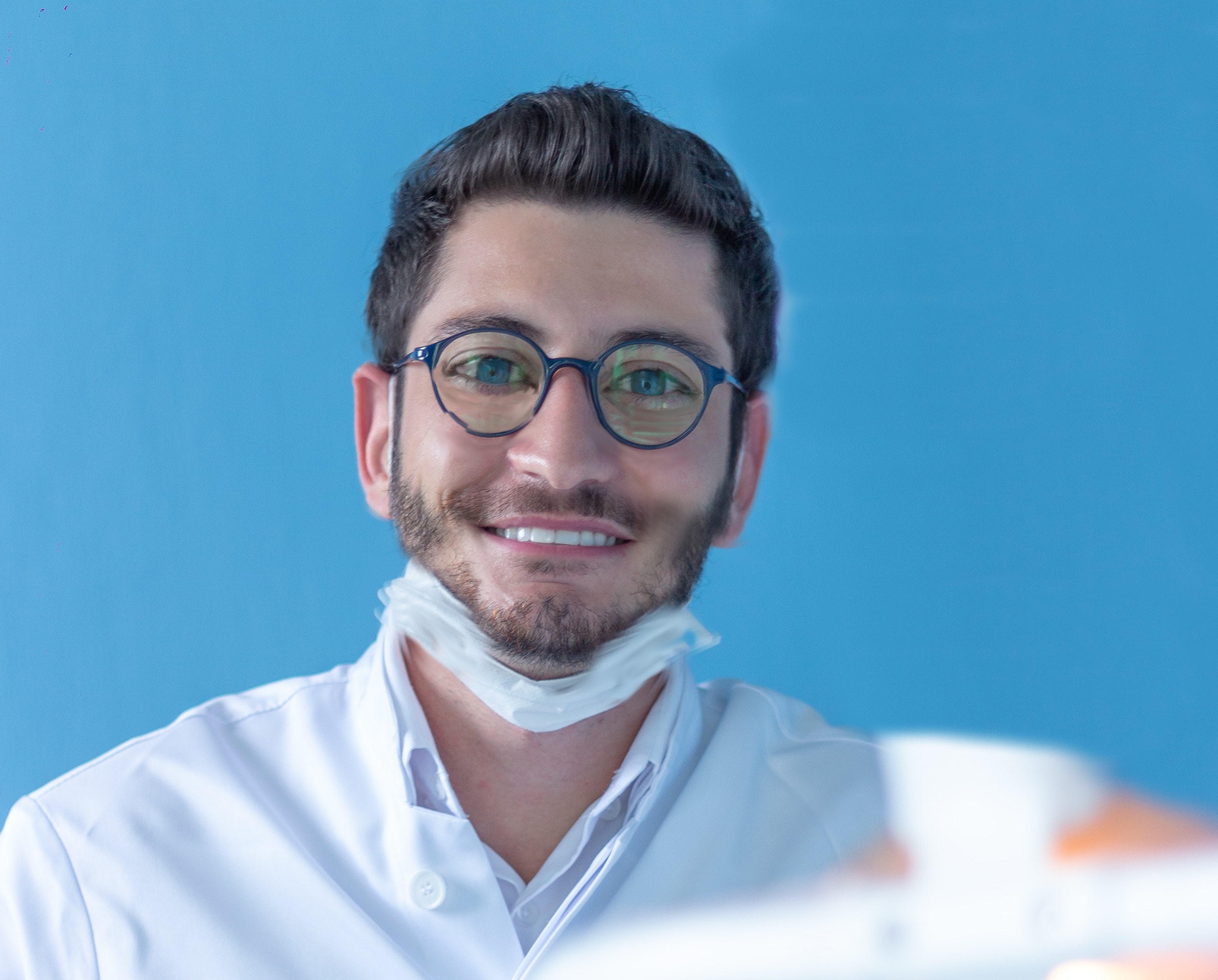}
  \caption{Image 2: Edited.}
  \label{doctorimage}
\end{subfigure}
\caption{Image Stimuli}
\label{doctorimages}
\end{figure}

Questions asked were as follows: While viewing, did you feel a sense of trust? (T), Did you feel that this person was attractive? (A), and Did you feel that this image may have been photoshopped? (M).
Question orders were $TAM$ and $TMA$ for each image.
Questions were selected based on the likelihood that the operators associated with these variables would be non-commutative. In other words, we were expecting some order effects between variables/operators, meaning that they are not entirely independent of one another. For example, we expect the probabilities associated with the question of attractiveness (‘A’, given ‘T’) and the probabilities associated with the question of manipulated (‘M’, given ‘T’ \& ‘A’) to be different if the order of ‘A’ and ‘M’ were to be reversed (i.e., $TMA$, with $M|T$, and $A|T$ \& $M$). The image stimuli used to gather ratings of the above dimensions are shown in Figure \ref{doctorimages}.
\subsection{Design}
Each image was presented with two question orders, creating a between subjects design with four conditions. The dependant variables were ratings of trustworthiness, attractiveness and image manipulation.

\begin{figure}
    \includegraphics[width=1.2\columnwidth,center]{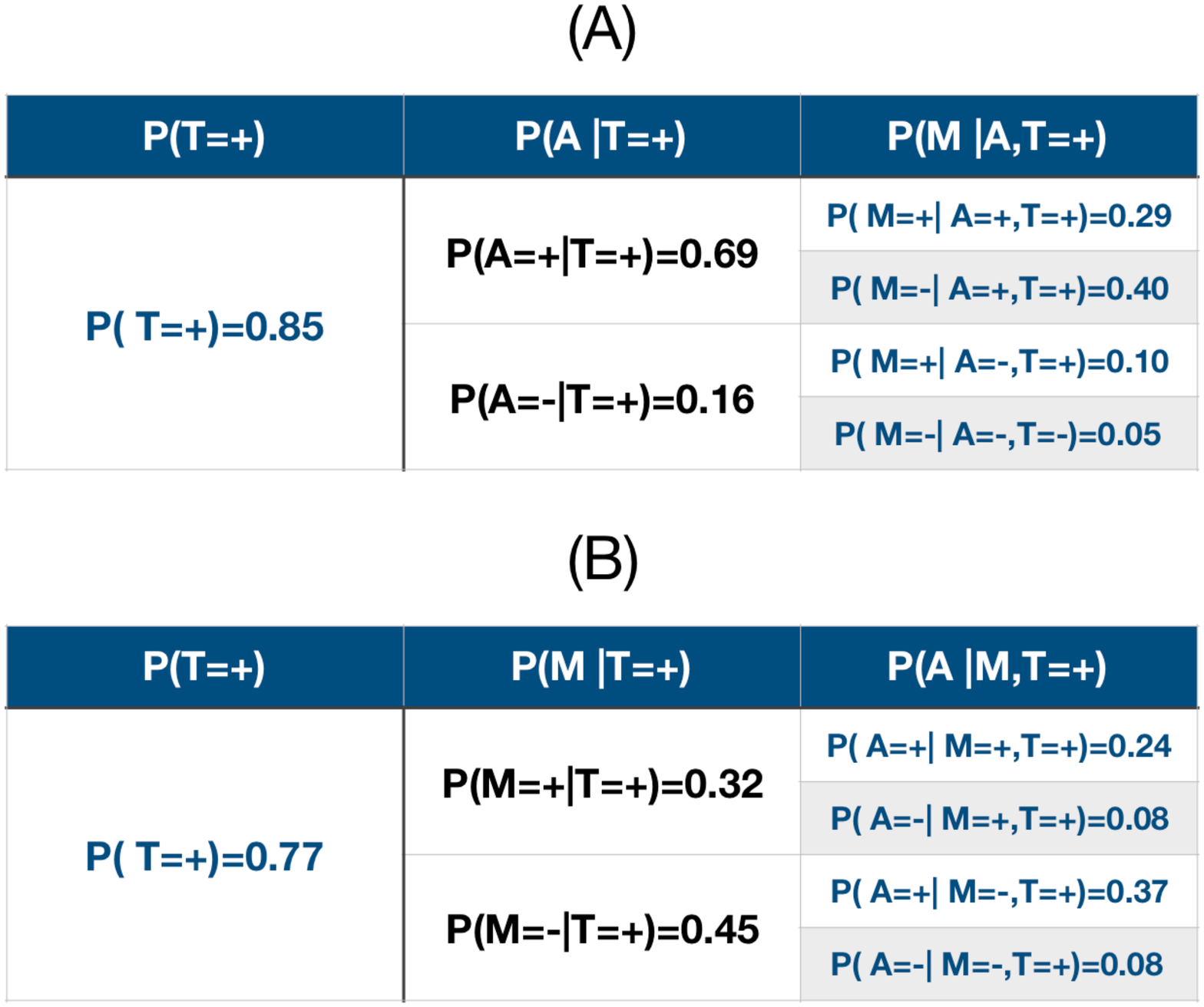}
    \caption{Table (A) and (B) correspond to Image 1 (unedited). Probabilities relating to the first question ($T$) are depicted in the first columns; the second column states  conditional probabilities of the second question given first one, i.e., $P(A=\pm|T=+)$ in Table (A) and $P(M=\pm|T=+)$ in Table (B); the third column indicates conditional probabilities of the third question given the first and second questions.}
    \label{fig5_mod}
\end{figure}

\begin{figure}[h]
    \centering
    \includegraphics[width=9cm]{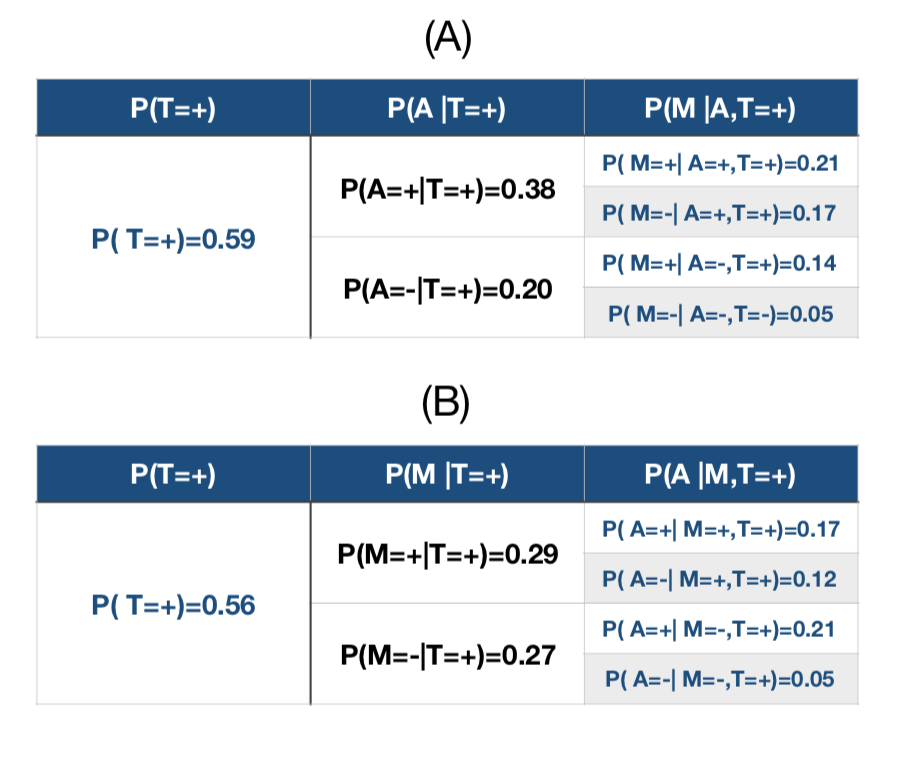}
    \caption{Table (A) and (B) correspond to Image 2 (edited). Probabilities relating to the first question ($T$) are depicted in the first columns; the second column states  conditional probabilities of the second question given first one,i.e., $P(A=\pm|T=+)$ in Table (A) and $P(M=\pm|T=+)$ in Table (B); the third column indicates conditional probabilities of the third question given the first and second questions.}
    \label{fig6_mod}
\end{figure}
\subsection{Procedure}
In all conditions, participants completed an online experiment by first perusing a short description on the Prolific site, if deciding to continue, they then clicked a link to the project page which begins with short instructions and a link to the consent form to read before continuing. The design of the experiment was aimed at accessing fast intuitive responses, rather than responses based on analytical thinking, as this was believed to be analogous to the short distances between measurement devices in the S.G. experiment (i.e. fast measurements restricting interacting influences). To this end, instructions included a notice to look out for a button popping up for some participants that afforded a bonus (distraction to assign less cognitive resources to the decision task), questions were asked with emotive wording (to help prompt intuitive thinking), and both image display and questions included a time limit (2 seconds for the image and 4 seconds for each question). Participants could only view one question at a time, with each subsequent question hidden until an answer had been given for the preceding one. Lastly, participants were asked to provide one or two words to describe their first impressions of what they saw, as well as a confidence rating for their combined judgments, and were asked their gender and the country they resided in. 
\subsection{Results}
\begin{figure*}
\vspace{-1.5 cm}
\begin{center}
    \includegraphics[width=18cm]{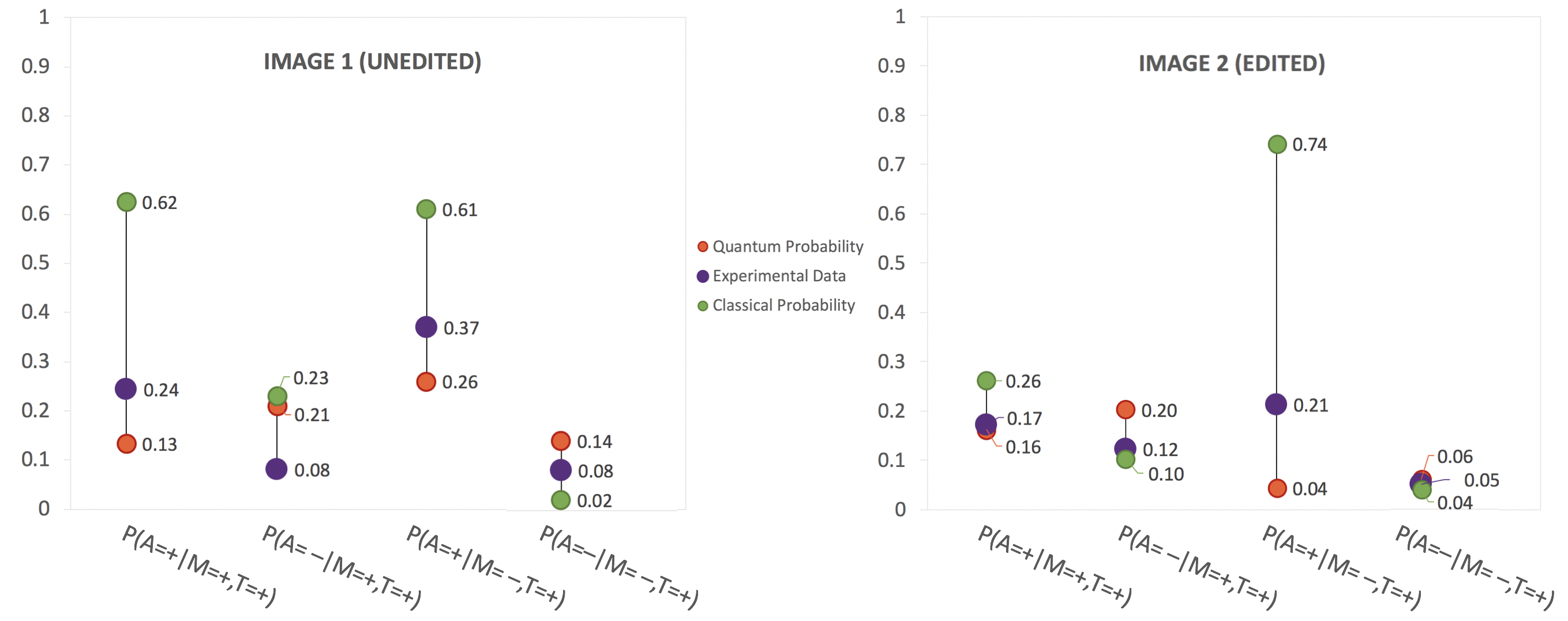}
    \caption{ Comparison of actual probabilities (Experimental Data) of the third question Attractiveness ($A$) to predicted probabilities computed by  the Born rule (Quantum Probabilities) and Bayes' rule (Classical Probabilities).
}\label{table3}
\end{center}
\end{figure*}
Based on the probabilities shown in Table \ref{fig5_mod} (A), (B) and Table \ref{fig6_mod} (A), (B), the cognitive states associated with the first question $T$ are given by:
\begin{eqnarray}
|T_{1}\rangle&=&\sqrt{0.85} |T_{1},+\rangle +\sqrt{0.15} |T_{1},-\rangle,\\
    |T_{2}\rangle&=&\sqrt{0.59} |T_{2},+\rangle+\sqrt{0.31} |T_{2},-\rangle.
\end{eqnarray}
where the subscripts respectively denote the unedited and edited images.\\
\indent 
According to probabilities in the second column in Table \ref{fig5_mod} (A), (B) and also by using the equation (\ref{eq3}), the angles between operator $\hat{T}$ and $\hat{A}$, as well as $\hat{T}$ and $\hat{M}$, are respectively given by
$\theta^{(1)}_{a}=51.42$ and $\theta^{(1)}_{m}=99.79$ for the unedited image. By using the same method, in the second column of Table \ref{fig6_mod} (A), (B), the angles between the operators $\hat{T}$ and $\hat{A}$, as well as $\hat{T}$ and $\hat{M}$, are obtained respectively by
$\theta^{(2)}_{a}=71.20$ and $\theta^{(2)}_{m}=87.70$ for the edited image. By using equation (\ref{eqf3}) together with the third column of Table \ref{fig5_mod} (A), (B) and Table \ref{fig6_mod} (A), (B), we obtain respectively $\phi^{(1)}_{m}=85.99$ and $\phi^{(2)}_{m}=85.30$ for the unedited and edited images.
\subsubsection*{Comparison of prediction of decision: quantum probabilities vs. classical probabilities}
For the prediction comparison, we consider a new situation in which the the order of questions is altered. According to operators $\hat{A}$ and $\hat{M}$  and new preparation state $|T\rangle$, we obtain the probability of positive and negative answers and compare them with the experimental data. 
Indeed, as defined in equation (\ref{eqf3}), the phase interference $\phi_{m}$ appears in the probability of the third question. 
The predictions of the quantum model are compared to classical probabilities in the following way:
The cognitive S.G. device depicted in Figure ~\ref{fig1} can be modelled by using the chain rule: $P(T,A,M) = P(T)P(A|T)P(M|A,T)$.
The three distributions on the RHS are empirically collected from the S.G. device.
Similarly, in the new situation the order of the $A$ and $M$ magnets are reversed so the chain rule is written out as follows: $P(T,M,A) = P(T)P(M|T)P(A|M,T)$. Therefore, 
\begin{eqnarray}
P(M|A,T) &= \frac{P(M|T)P(A|M,T)}{P(A|T)} \\
P(A|M,T) &= \frac{P(A|T)P(M|A,T)}{P(M|T)}
\end{eqnarray}
The LHS of both equations constitute predictions based on classical probability theory.
As evidenced by Figure \ref{table3} (A) and (B), the predicted results calculated based on the HSM are generally closer to the actual probabilities than the classical predictions. 
Figure \ref{table3} compares results of probabilities of the decision regarding manipulation given attractiveness and trustworthiness, for the unedited and edited images respectively. 
\\
\subsubsection{Wigner functions for both images}
 By using equation (\ref{eq7}):
\begin{eqnarray}
 r_{x}=2\sqrt{P_{t}(+)\left(1-P_{t}(+)\right)},\ 
 r_{y}=0,\ 
r_{z} =2P_{t}(+)-1, \nonumber
\end{eqnarray}
The discrete Wigner distribution that is obtained is the following:
\begin{eqnarray}
W=\frac{1}{4}\left(
\begin{array}{cc}
1+r_{x}+r_{z}     &  1-r_{x}+r_{z} \\
1-r_{x}-r_{z}     & 1+r_{x}-r_{z}
\end{array}
\right)
\end{eqnarray}
Therefore, the Wigner distributions for both  unedited ($W_{1}$) and edited ($W_{2}$) images are given as follows:
\begin{eqnarray}
W_{1}=\left(
\begin{array}{cc}
0.63     & 0.13 \\
-0.13     & 0.36
\end{array}
\right),
\hspace{.5cm}
W_{2}=\left(
\begin{array}{cc}
0.53     & 0.03 \\
-0.03     & 0.47
\end{array}
\right)
\end{eqnarray}

\subsection{Discussion}
The Wigner function of both images showed negative values. 
Therefore, the cognitive analogue of the S.G. experiment that produces quantum models in physics, also produces a quantum model for cognitive decision making.\\
\indent It is known from physics that negative values in the Wigner function are a consequence of quantum interference effects.
The negative values are a consequence of the fact that once a particle has passed through a magnet its polarization (either + or -) is not retained when it arrives at the next magnet. 
This is a consequence of the the fact that a particle is always in a superposed state each time it interacts with a magnet. As a result of the interaction, a particular polarization will be observed.  
In terms of the cognitive analogue depicted in Figure ~\ref{fig1}, the preceding can be translated as follows: Even though a subject has already decided that they trust (T=+) the image and have deemed the face to be attractive (A=+), when they are presented with the decision about whether the image is manipulated, at that decision point they are necessarily superposed with respect to trust and attractiveness.
This can only occur when the decision perspectives are incompatible.
Incompatibility is indeed present in the HSM as the operators corresponding to decisions of trustworthiness $\hat{T},\hat{A},\hat{M}$ do not pair-wise mutually commute: $[\hat{T},\hat{A}] \neq 0$, $[\hat{T},\hat{M}] \neq 0$, $[\hat{A},\hat{M}] \neq 0$.\\
\indent Incompatibility generates interference effects which generate probabilities of outcomes, that is they are fundamentally different from standard probabilistic models \cite{bruza:wang:busemeyer:2015}.
As stated above, the cognitive S.G. device depicted in Figure ~\ref{fig1} can be modelled by using the chain rule: $P(T,A,M) = P(T)P(A|T)P(M|A,T)$. This expresses that the underlying probabilistic model of the device is simply the joint probability distribution $P(T,A,M)$. 
The critical point is that  the structure of the  event space underpinning $P(T,A,M)$ assumes that the variables are \emph{jointly} measurable, e.g., the subject can simultaneously access information regarding the attractiveness of the face and whether the image is manipulated. The previously mentioned incompatibility in the HSM  $[\hat{A},\hat{M}] \neq 0$ implies that this assumption does not hold. Consequently, the subject cannot cognitively form the joint distribution $P(T,A,M)$.
In short, the HSM provides a probabilistic framework which does not rely on the assumption that variables are jointly measurable.
This has been one of the key features of quantum models of cognition \cite{busemeyer:bruza:2012}.\\
\indent
The use of three operators is crucial in the derivation of the two-dimensional Hilbert space because three operators necessarily entail that a {\it complex} Hilbert space must be used. The use of less than three operators necessarily implies that the cognitive decision model can be expressed as a real-valued Hilbert space, which has been the practice thus far in quantum cognition research. 
The significance of this difference lies in the complex phase factor $\exp(i\phi_m)$ which cannot be derived unless there are three operators. We speculate that it is this  phase factor which generates the interference effects for the Wigner function to go negative and hence become quantum. 
To the best of our knowledge, this study is the first to: a) develop a specialised protocol to genuinely exploit the complex Hilbert space by constructing three operators and states, and b) utilise the Wigner function to determine the quantumness of a cognitive state. \\
\indent 
Moreover, this determination is straightforward and does not suffer from the challenges and controversies associated with using contextuality to determine whether the cognitive system is quantum-like \cite{dzhafarov:kujala:cervantes:zhang:jones:2016,bruza:fell:2018}.
\section{Conclusions and future work}
This article has demonstrated the specification and validation of a quantum decision model by employing an experimental protocol derived from quantum physics.
The protocol involved three  binary decisions in a forced choice design. Future studies may investigate the measurement of decisions by asking binary questions of any number of points within a spectrum of responses by extending the quantum model described in this paper. 
\section{Acknowledgments}
This research was supported by the Asian Office of Aerospace Research and Development (AOARD) grant: FA2386-17-1-4016

\nocite{ChalnickBillman1988a}
\nocite{Feigenbaum1963a}
\nocite{Hill1983a}
\nocite{OhlssonLangley1985a}
\nocite{Matlock2001}
\nocite{NewellSimon1972a}
\nocite{ShragerLangley1990a}

\bibliographystyle{apacite}

\setlength{\bibleftmargin}{.125in}
\setlength{\bibindent}{-\bibleftmargin}


\begin{thebibliography}{}

\bibitem [\protect \citeauthoryear {%
Bruza%
\ \BBA {} Fell%
}{%
Bruza%
\ \BBA {} Fell%
}{%
{\protect \APACyear {2018}}%
}]{%
bruza:fell:2018}
\APACinsertmetastar {%
bruza:fell:2018}%
\begin{APACrefauthors}%
Bruza, P.%
\BCBT {}\ \BBA {} Fell, L.%
\end{APACrefauthors}%
\unskip\
\newblock
\APACrefYearMonthDay{2018}{}{}.
\newblock
{\BBOQ}\APACrefatitle {Are decisions of image trustworthiness contextual? A
  pilot study} {Are decisions of image trustworthiness contextual? a pilot
  study}.{\BBCQ}
\newblock
\BIn{} A.~Lambert-Mogiliansky\ \BBA {} B.~Coecke\ (\BEDS), \APACrefbtitle
  {Quantum Interaction: 11th International Conference (QI'2018).} {Quantum
  interaction: 11th international conference (qi'2018).}
\newblock
\APACaddressPublisher{}{Springer}.
\PrintBackRefs{\CurrentBib}

\bibitem [\protect \citeauthoryear {%
Bruza%
, Wang%
\BCBL {}\ \BBA {} Busemeyer%
}{%
Bruza%
\ \protect \BOthers {.}}{%
{\protect \APACyear {2015}}%
}]{%
bruza:wang:busemeyer:2015}
\APACinsertmetastar {%
bruza:wang:busemeyer:2015}%
\begin{APACrefauthors}%
Bruza, P.%
, Wang, Z.%
\BCBL {}\ \BBA {} Busemeyer, J\BPBI R.%
\end{APACrefauthors}%
\unskip\
\newblock
\APACrefYearMonthDay{2015}{}{}.
\newblock
{\BBOQ}\APACrefatitle {Quantum cognition: a new theoretical approach to
  psychology} {Quantum cognition: a new theoretical approach to
  psychology}.{\BBCQ}
\newblock
\APACjournalVolNumPages{Trends in Cognitive Sciences}{19}{7}{383 - 393}.
\PrintBackRefs{\CurrentBib}

\bibitem [\protect \citeauthoryear {%
Busemeyer%
\ \BBA {} Bruza%
}{%
Busemeyer%
\ \BBA {} Bruza%
}{%
{\protect \APACyear {2012}}%
}]{%
busemeyer:bruza:2012}
\APACinsertmetastar {%
busemeyer:bruza:2012}%
\begin{APACrefauthors}%
Busemeyer, J.%
\BCBT {}\ \BBA {} Bruza, P.%
\end{APACrefauthors}%
\unskip\
\newblock
\APACrefYear{2012}.
\newblock
\APACrefbtitle {Quantum cognition and decision} {Quantum cognition and
  decision}.
\newblock
\APACaddressPublisher{}{Cambridge University Press}.
\PrintBackRefs{\CurrentBib}

\bibitem [\protect \citeauthoryear {%
Di~Matteo%
, S\'anchez-Soto%
, Leuchs%
\BCBL {}\ \BBA {} Grassl%
}{%
Di~Matteo%
\ \protect \BOthers {.}}{%
{\protect \APACyear {2017}}%
}]{%
PhysRevA.95.022340}
\APACinsertmetastar {%
PhysRevA.95.022340}%
\begin{APACrefauthors}%
Di~Matteo, O.%
, S\'anchez-Soto, L\BPBI L.%
, Leuchs, G.%
\BCBL {}\ \BBA {} Grassl, M.%
\end{APACrefauthors}%
\unskip\
\newblock
\APACrefYearMonthDay{2017}{Feb}{}.
\newblock
{\BBOQ}\APACrefatitle {Coarse graining the phase space of $N$ qubits} {Coarse
  graining the phase space of $n$ qubits}.{\BBCQ}
\newblock
\APACjournalVolNumPages{Phys. Rev. A}{95}{}{022340}.
\PrintBackRefs{\CurrentBib}

\bibitem [\protect \citeauthoryear {%
Dzhafarov%
, Kujala%
, Cervantes%
, Zhang%
\BCBL {}\ \BBA {} Jones%
}{%
Dzhafarov%
\ \protect \BOthers {.}}{%
{\protect \APACyear {2016}}%
}]{%
dzhafarov:kujala:cervantes:zhang:jones:2016}
\APACinsertmetastar {%
dzhafarov:kujala:cervantes:zhang:jones:2016}%
\begin{APACrefauthors}%
Dzhafarov, E.%
, Kujala, J.%
, Cervantes, V.%
, Zhang, R.%
\BCBL {}\ \BBA {} Jones, M.%
\end{APACrefauthors}%
\unskip\
\newblock
\APACrefYearMonthDay{2016}{}{}.
\newblock
{\BBOQ}\APACrefatitle {On contextuality in behavioural data} {On contextuality
  in behavioural data}.{\BBCQ}
\newblock
\APACjournalVolNumPages{Philosophical Transactions of the Royal Society A:
  Mathematical, Physical and Engineering Sciences}{374}{}{}.
\PrintBackRefs{\CurrentBib}

\bibitem [\protect \citeauthoryear {%
Galvao%
}{%
Galvao%
}{%
{\protect \APACyear {2005}}%
}]{%
galvao2005discrete}
\APACinsertmetastar {%
galvao2005discrete}%
\begin{APACrefauthors}%
Galvao, E\BPBI F.%
\end{APACrefauthors}%
\unskip\
\newblock
\APACrefYearMonthDay{2005}{}{}.
\newblock
{\BBOQ}\APACrefatitle {Discrete Wigner functions and quantum computational
  speedup} {Discrete wigner functions and quantum computational
  speedup}.{\BBCQ}
\newblock
\APACjournalVolNumPages{Physical Review A}{71}{4}{042302}.
\PrintBackRefs{\CurrentBib}

\bibitem [\protect \citeauthoryear {%
Gibbons%
, Hoffman%
\BCBL {}\ \BBA {} Wootters%
}{%
Gibbons%
\ \protect \BOthers {.}}{%
{\protect \APACyear {2004}}%
}]{%
gibbons2004discrete}
\APACinsertmetastar {%
gibbons2004discrete}%
\begin{APACrefauthors}%
Gibbons, K\BPBI S.%
, Hoffman, M\BPBI J.%
\BCBL {}\ \BBA {} Wootters, W\BPBI K.%
\end{APACrefauthors}%
\unskip\
\newblock
\APACrefYearMonthDay{2004}{}{}.
\newblock
{\BBOQ}\APACrefatitle {Discrete phase space based on finite fields} {Discrete
  phase space based on finite fields}.{\BBCQ}
\newblock
\APACjournalVolNumPages{Physical Review A}{70}{6}{062101}.
\PrintBackRefs{\CurrentBib}

\bibitem [\protect \citeauthoryear {%
Goh%
\ \protect \BOthers {.}}{%
Goh%
\ \protect \BOthers {.}}{%
{\protect \APACyear {2018}}%
}]{%
goh2018geometry}
\APACinsertmetastar {%
goh2018geometry}%
\begin{APACrefauthors}%
Goh, K\BPBI T.%
, Kaniewski, J.%
, Wolfe, E.%
, V{\'e}rtesi, T.%
, Wu, X.%
, Cai, Y.%
\BDBL {}Scarani, V.%
\end{APACrefauthors}%
\unskip\
\newblock
\APACrefYearMonthDay{2018}{}{}.
\newblock
{\BBOQ}\APACrefatitle {Geometry of the set of quantum correlations} {Geometry
  of the set of quantum correlations}.{\BBCQ}
\newblock
\APACjournalVolNumPages{Physical Review A}{97}{2}{022104}.
\PrintBackRefs{\CurrentBib}

\bibitem [\protect \citeauthoryear {%
Huang%
, Yu%
\BCBL {}\ \BBA {} Zhang%
}{%
Huang%
\ \protect \BOthers {.}}{%
{\protect \APACyear {{\protect \bibnodate {}}}}%
}]{%
huangnegativity}
\APACinsertmetastar {%
huangnegativity}%
\begin{APACrefauthors}%
Huang, M\BHBI D.%
, Yu, Y\BHBI F.%
\BCBL {}\ \BBA {} Zhang, Z\BHBI M.%
\end{APACrefauthors}%
\unskip\
\newblock
\APACrefYearMonthDay{{\protect \bibnodate {}}}{}{}.
\newblock
{\BBOQ}\APACrefatitle {The Negativity-to-Violation Map between Wigner Function
  and Quantum Contextuality Inequality for a Single Qudit} {The
  negativity-to-violation map between wigner function and quantum contextuality
  inequality for a single qudit}.{\BBCQ}
\newblock
\APACjournalVolNumPages{Annalen der Physik}{}{}{1800464}.
\PrintBackRefs{\CurrentBib}

\bibitem [\protect \citeauthoryear {%
Kocia%
\ \BBA {} Love%
}{%
Kocia%
\ \BBA {} Love%
}{%
{\protect \APACyear {2017}}%
}]{%
kocia2017discrete}
\APACinsertmetastar {%
kocia2017discrete}%
\begin{APACrefauthors}%
Kocia, L.%
\BCBT {}\ \BBA {} Love, P.%
\end{APACrefauthors}%
\unskip\
\newblock
\APACrefYearMonthDay{2017}{}{}.
\newblock
{\BBOQ}\APACrefatitle {Discrete Wigner formalism for qubits and
  noncontextuality of Clifford gates on qubit stabilizer states} {Discrete
  wigner formalism for qubits and noncontextuality of clifford gates on qubit
  stabilizer states}.{\BBCQ}
\newblock
\APACjournalVolNumPages{Physical Review A}{96}{6}{062134}.
\PrintBackRefs{\CurrentBib}

\bibitem [\protect \citeauthoryear {%
Raussendorf%
, Browne%
, Delfosse%
, Okay%
\BCBL {}\ \BBA {} Bermejo-Vega%
}{%
Raussendorf%
\ \protect \BOthers {.}}{%
{\protect \APACyear {2017}}%
}]{%
raussendorf2017contextuality}
\APACinsertmetastar {%
raussendorf2017contextuality}%
\begin{APACrefauthors}%
Raussendorf, R.%
, Browne, D\BPBI E.%
, Delfosse, N.%
, Okay, C.%
\BCBL {}\ \BBA {} Bermejo-Vega, J.%
\end{APACrefauthors}%
\unskip\
\newblock
\APACrefYearMonthDay{2017}{}{}.
\newblock
{\BBOQ}\APACrefatitle {Contextuality and Wigner-function negativity in qubit
  quantum computation} {Contextuality and wigner-function negativity in qubit
  quantum computation}.{\BBCQ}
\newblock
\APACjournalVolNumPages{Physical Review A}{95}{5}{052334}.
\PrintBackRefs{\CurrentBib}

\bibitem [\protect \citeauthoryear {%
Sakurai%
\ \BBA {} Commins%
}{%
Sakurai%
\ \BBA {} Commins%
}{%
{\protect \APACyear {1995}}%
}]{%
sakurai1995modern}
\APACinsertmetastar {%
sakurai1995modern}%
\begin{APACrefauthors}%
Sakurai, J\BPBI J.%
\BCBT {}\ \BBA {} Commins, E\BPBI D.%
\end{APACrefauthors}%
\unskip\
\newblock
\APACrefYearMonthDay{1995}{}{}.
\newblock
\APACrefbtitle {Modern quantum mechanics, revised edition.} {Modern quantum
  mechanics, revised edition.}
\newblock
\APACaddressPublisher{}{AAPT}.
\PrintBackRefs{\CurrentBib}

\bibitem [\protect \citeauthoryear {%
Tversky%
\ \BBA {} Kahneman%
}{%
Tversky%
\ \BBA {} Kahneman%
}{%
{\protect \APACyear {1974}}%
}]{%
tversky1974judgment}
\APACinsertmetastar {%
tversky1974judgment}%
\begin{APACrefauthors}%
Tversky, A.%
\BCBT {}\ \BBA {} Kahneman, D.%
\end{APACrefauthors}%
\unskip\
\newblock
\APACrefYearMonthDay{1974}{}{}.
\newblock
{\BBOQ}\APACrefatitle {Judgment under uncertainty: Heuristics and biases}
  {Judgment under uncertainty: Heuristics and biases}.{\BBCQ}
\newblock
\APACjournalVolNumPages{science}{185}{4157}{1124--1131}.
\PrintBackRefs{\CurrentBib}

\bibitem [\protect \citeauthoryear {%
Vourdas%
}{%
Vourdas%
}{%
{\protect \APACyear {2019}}%
}]{%
vourdas2019probabilistic}
\APACinsertmetastar {%
vourdas2019probabilistic}%
\begin{APACrefauthors}%
Vourdas, A.%
\end{APACrefauthors}%
\unskip\
\newblock
\APACrefYearMonthDay{2019}{}{}.
\newblock
{\BBOQ}\APACrefatitle {Probabilistic inequalities and measurements in bipartite
  systems} {Probabilistic inequalities and measurements in bipartite
  systems}.{\BBCQ}
\newblock
\APACjournalVolNumPages{Journal of Physics A: Mathematical and
  Theoretical}{}{}{}.
\PrintBackRefs{\CurrentBib}

\bibitem [\protect \citeauthoryear {%
Wootters%
}{%
Wootters%
}{%
{\protect \APACyear {1987}}%
}]{%
wootters1987wigner}
\APACinsertmetastar {%
wootters1987wigner}%
\begin{APACrefauthors}%
Wootters, W\BPBI K.%
\end{APACrefauthors}%
\unskip\
\newblock
\APACrefYearMonthDay{1987}{}{}.
\newblock
{\BBOQ}\APACrefatitle {A Wigner-function formulation of finite-state quantum
  mechanics} {A wigner-function formulation of finite-state quantum
  mechanics}.{\BBCQ}
\newblock
\APACjournalVolNumPages{Annals of Physics}{176}{1}{1--21}.
\PrintBackRefs{\CurrentBib}

\end{thebibliography}

\end{document}